# Strategic improvement of second and third harmonic generation in multifunctional Cu-Sn-S$_3$ ternary semiconducting thin films

B. Sudarshan Acharya[a], Anushaa, Albin Antony[a], Aninamol Ani[a], I.V. Kityk[b,*], K.Ozga[b], A. Slezak[c], J. Jedryka[b], P. Poornesh[a,*], K.B. Manjunatha[d], Shashidhara Acharya[e] a Department of Physics, Manipal Institute of Technology, Manipal Academy of Higher Education, Manipal, Karnataka 576104, India
b Institute of Optoelectronics and Measuring Systems, Faculty of Electrical Engineering, Czestochowa University of Technology, Armii Krajowej 17, PL-42-201 Czestochowa, Poland
c Faculty of Management, Czestochowa Univeristy of Technology, Armii Krajowej,19 B, PL-42201, Czestochowa, Poland
d Department of Physics, NMAM Institute of Technology, Nitte 574110, India
e International Centre for Materials Science,Jawaharlal Nehru Centre for Advanced Scientific Research, Jakkur, Bangalore 64, India

ABSTRACT

We propose a low-cost approach for the synthesis of multifunctional CuSnS$_3$ (CTS) ternary compound thin film via spray pyrolysis technique. By varying Sn and S doping concentrations, a high energy absorber layers of CuSnS$_3$ thin films were deposited on a glass substrate at a substrate temperature of 405 °C. The prepared samples were analysed with respect to their optical, structural and electrical and nonlinear optical properties. X-Ray diffraction (XRD) analysis reveals that the films exhibit a tetragonal crystal structure with a preferential growth orientation along (1 1 2). The surface morphology of the films was explored by atomic force microscopy (AFM) in tapping mode configuration. Variation in the carrier charge density and electrical properties were observed for different Sn and S combination. The analysis of the Raman spectra indicates the presence of multiple phases apart from CuSnS$_3$. The obtained Raman spectra were assigned to phonon mode as per zone centre phonon representation of optical and acoustic modes and identified to dominant "A" symmetry modes. The maximal laser stimulated induced second harmonic generation (SHG) signal was observed for the CTS 3 film and the corresponding second order nonlinear optical susceptibility which was equal to about 0.89 pm/V at 1064 nm and the minimal SHG signal was found for the CTS 2 film (about 0.22 pm/V). This strategic improvement in SHG and third harmonic generation (THG) signal efficiency endorses the role of Sn and S in modulating second and third harmonic generations in CuSnS3 compound.

1. Introduction

Semiconducting nanostructures are of considerable interest due to their unique optical and nonlinear optical (NLO) properties. Additional interest of their studies consists in the further investigation of new NLO materials with modified features in the optoelectronics industry to meet the present demand [1].Second harmonic generation (SHG) and third harmonic generation (THG) are the well-known nonlinear optical phenomena's. SHG and THG activated materials with high conversion efficiency have wide application in the field of photonics and laser technology as frequency doublers and tipplers. CuSnS$_3$ is ternary, p-type semiconductor compound which possesses a well-explored photovoltaic absorber layers with a direct band gap varied within 1.4–1.6 eV [2–6]. A lot of investigations have been carried out on this material in this context. Moreover, a higher order of absorption exhibited by this ternary compound can be utilized for a wide range of NLO device applications. However, in-depth understanding of NLO properties of CTS compound yet to be explored further in order to utilize this material for multifunctional nonlinear optical device applications [2]. Most of the previous works on the films were devoted to modification of their structure using doping or amending a chemical technology [7–10] In the present work, we try to use a novel strategy for improvement of nonlinear optical susceptibilities using both photoinduced continuous wave (CW) external laser light beams together with appropriate doping of the films. The basic principle in our approach presented here is grounded on the occurrence of the internal local dc-electric field due to the interaction of the external CW laser field coherent beams with the internal charged defect states occurred due to the doping [11,12] For the case of the chalcogenide semiconducting films a crucial role is played also by enhanced phonon anharmonicities which favour an additional increase of the corresponding laser stimulated effects [13]. Such an approach may be extended for other crystalline films possessing high phonon anharmonicities during photo treatment due to their NLO features in the desirable direction. Apart from the inherent NLO properties of the materials, the quality of films is an important parameter for NLO device applications. Preparation method and parameters are important to get the desired quality of thin crystalline films. Many methods are used in preparing of thin films, for instance chemical deposition, solution growth, screen printing, physical vapour deposition technique etc… These methods are time-consuming, expensive and not suitable for large scale production. Spray pyrolysis technique is a solution based method useful for preparing good quality thin films with less time and low cost [14]. In the present study, we implemented a spray pyrolysis technique and successfully prepared $CuSnS_3$ thin film samples with different combinations of Tin (Sn) and Sulphur (S) and the role of Sn and S concentration in the enhancement of second and third harmonic efficiencies were demon-strated and reported.

2. Experimental details

Using a spray pyrolysis technique $CuSnS_3$ crystalline thin films were deposited on SLG (soda lime glass) substrate successfully. An aqueous solution of content 0.05 M $CuCl_2 \cdot 2H_2O$, 0.05 M $SnCl_4 \cdot 5H_2O$ and 0.075 M $CH_4N_2S \cdot 7H_2O$ was used [4]. The three different samples were prepared with different concentrations of Sn and S with a ratio of 10:30:60, 10:40:50, and 10:50:40 and was named as CTS 1, CTS 2 and CTS 3. The aqueous solution was sprayed by keeping SLG at 410 °C. The thickness of the prepared films was determined by the Stylus profilometer and found to be around ~350 nm. Compressed air under the pressure of 2 Pa was used as a carrier gas for spray pyrolysis technique. The spray nozzle was fixed at a distance of 28 cm above the substrates and the deposition rate was fixed to 2 ml/min. X-ray Diffraction with Cu Kα radiation, λ = 0.15406 nm was applied to determine the crystallinity, crystallite size, dislocation density and strain in the film. Atomic force microscopy (AFM) technique in tapping operation mode was used to analyse the surface morphological characteristics. The average surface roughness of the films was also estimated using nanoscope analysis software. Ambient temperature electrical resistivity and charge carrier density of the films were calculated using Hall Effect measurement technique. Analysis of structural properties of the films was extended using Raman spectroscopy analysis. The laser stimulated experiment was done using a continuous wave (CW) laser source at an excitation wavelength of 532 nm and power varying within the 100– 200 mJ with diameter varying within 7–11 mm. Harmonic generation in CTS films was investigated using fundamental 8 ns pulsed laser beam at a wavelengthof 1064 nm with pulse frequency repetition equal to about 12 HZ.

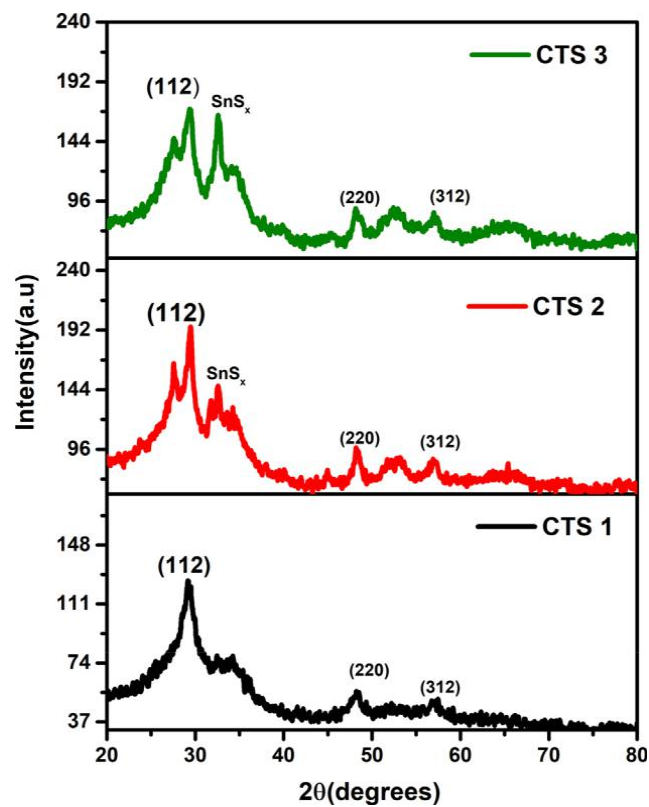

Fig. 1. XRD pattern of thetitled CTS thin films.

3. Result and discussions

3.1. Basic structural properties

The XRD patterns of CTS films are shown in Fig. 1.The films exhibit multiple growth phases with a preferential growth orientation along (1 1 2) plane. The obtained patterns have been indexed to JCPDS 01-089-4714 and have shown a tetragonal crystal structure [15–17]. The higher values of FWHM confirms the nanocrystalline origin of CTS films [15]. A characteristic peak of $SnS_x$ observed in the pattern for CTS 1 and CTS 2 combinations were due to the higher content of Sn in the CTS matrix [16].

The crystalline sizes of the films were calculated using Debye Scherer formula given by Eq. (1) and depicted in Table 1

$$d = \frac{K \cdot \lambda}{\beta \cdot cos\theta} \quad (1)$$

where K is a shape constant (k = 0.9), λ is the wavelength of the X-rays (0.154 nm for Cu-Kα), β is the full width at half maximum.

From the Table 1 it is noteworthy that structural parameters of the films do not show a substantial variation upon altering Sn: S content ratio in the matrix. The highest crystalline sizes with less dislocation density and lattice strain were observed for CTS 2 film with a $CuSnS_3$ content (10:40:50).

Fig. 2 shows the Raman spectra of the CTS thin crystalline films. The Raman peaks observed at frequencies 334 $cm^{-1}$ and 324 $cm^{-1}$ confirm the tetragonal phase of the deposited film. The presence of an extra peak around ~277 $cm^{-1}$ confirms the presence of secondary phases like $Cu_{2-x}S$ [15,3]. In the present study, the grown CTS films were assigned to tetragonal crystal structure as confirmed by XRD and Raman spec-troscopy. The observed Raman peaks can be attributed to various phonon modes occurring in the material as per the zone centre phonon representation of optical and acoustic modes [15,16]. The CTS films attributed to space group I-42 m (No. 121), point group D2d (−42 m)

Table 1
The basic structural parameters of the CTS thin films.

| Sample | 2θ (1 1 2) (deg) | FWHM (rad) | Crystalline Sized (nm) | Dislocation density δ ($m^{-2}$) × $10^{16}$ | Crystallites Per unit volume N ($m^{-3}$) × $10^{10}$ | Strain ε × $10^{-3}$ |
|---|---|---|---|---|---|---|
| CTS1 | 29.21 | 1.86 | 4.40 | 5.14 | 1.54 | 7.86 |
| CTS2 | 29.44 | 1.21 | 6.75 | 2.18 | 6.56 | 5.12 |
| CTS3 | 29.41 | 1.60 | 5.11 | 3.82 | 1.14 | 6.78 |

and contains one formula group per unit cell. Therefore according to Debye's theory of lattice vibration, there exist totally 36 vibrational modes which can be represented as [15]

M= 2A1 + A2 + 3B1 + 7B2 + 10E     (2)

The spectra of CTS films are depicted in Fig. 2 and were observed with two peaks which are identified to dominant "A" symmetry modes. A slight spectral shift observed in the phonon mode position was due to the variation in the composition of Sn and S in the matrix and also due to possible compressive strain in the CTS layer.

3.2. Surface morphology

Surface morphology of CTS films was studied by atomic force microscopy in tapping mode configuration. The root means square (RMS) surface roughness and surface morphology of the films were estimated and recorded using nanoscope analysis software which defines the quality of deposited films. From Fig. 3 it is observed that grains are densely packed with well-defined tetragonal shape and surfaces of samples are free of pinholes which result in a reduction of shunting problems during electrical property analysis [6]. The reason for the formation of well-defined CTS films was due to the high synthesis temperature of 405 °C which will prolong the time of sulfuration annealing and more Sn elements will evaporate in the form of SnS.[16] The sublimated SnS can promote the growth of grain and the film. Variation in RMS surface roughness was calculated using nanoscope analysis software and depicted in Table 2.

3.3. Optical properties

Optical absorbance of CTS thin films was investigated by employing UV visible absorption spectroscopy with specral resolution about 1 nm. The films show a saturated absorption in the visible spectrum within 420 nm 580 nm wavelengths. The CTS 3 films show maximum absorption value equal to 3.8 at wavelength about 350 nm. From the Fig. 4 it is observed that the thin film possessing a saturated region which is higher for CTS 3 film. This inherent high absorbing property of CTS thin film can be widely explored as an absorber layer for photovoltaic applications. From the absorbance values, the energy band gap of the films was estimated by Tauc plot analysis as shown in Fig. 5. The energy band gap shows an enhancement and was evaluated to be around 1.50 eV, 1.65 eV and 1.76 which is in good agreement with the reported values [15–17].

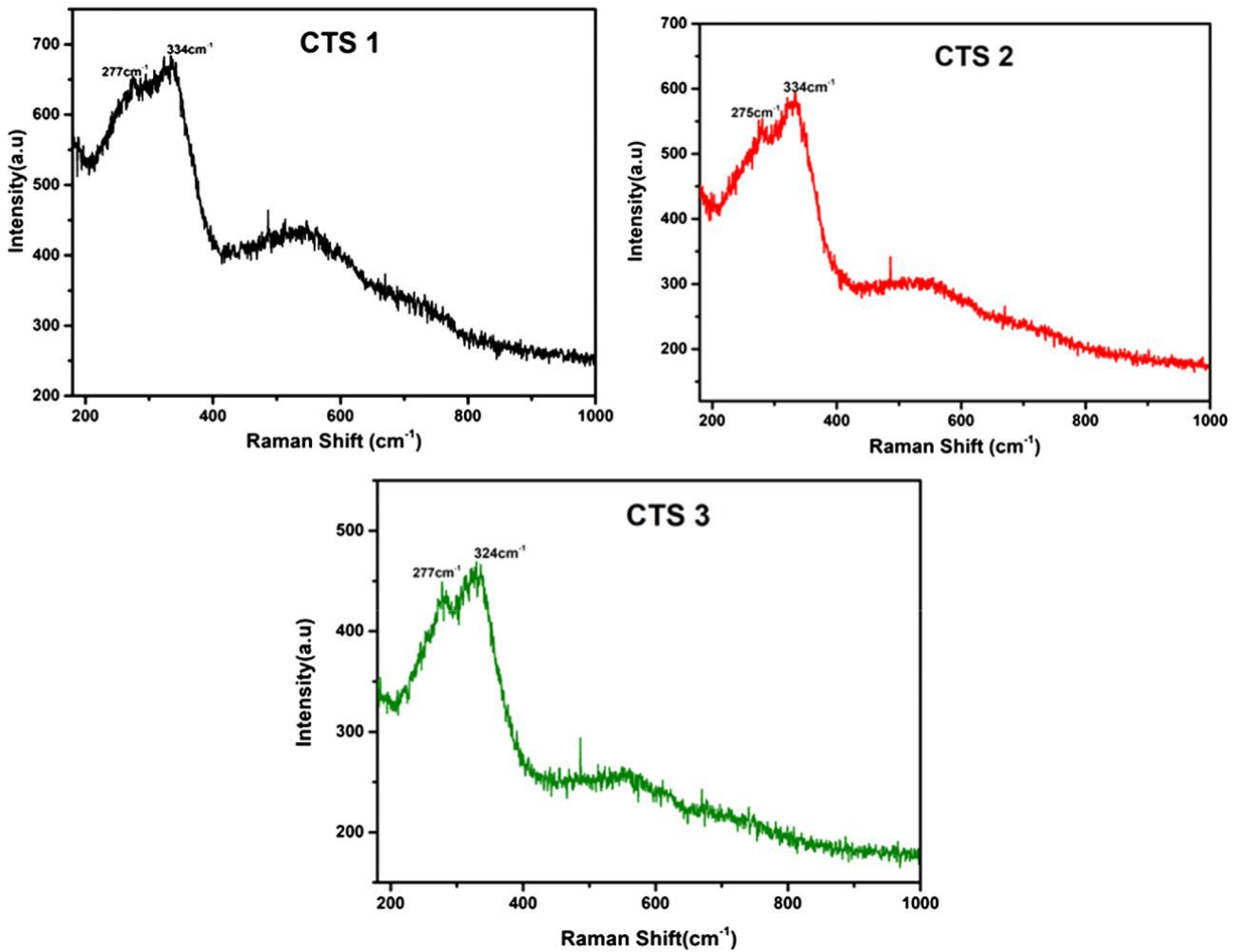

Fig. 2. Raman Spectra of the studied CTS thin films.

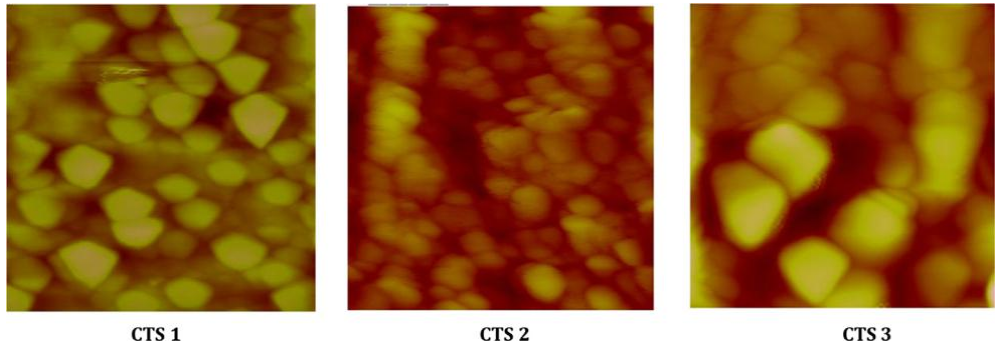

Fig. 3. 3D AFM Images of CTS thin films.

Table 2
RMS surface roughness variation upon different Sn and S combinations.

| Sample | RMS Roughness (nm) |
| --- | --- |
| CTS 1 | 23.1 |
| CTS 2 | 14.6 |
| CTS 3 | 36.0 |

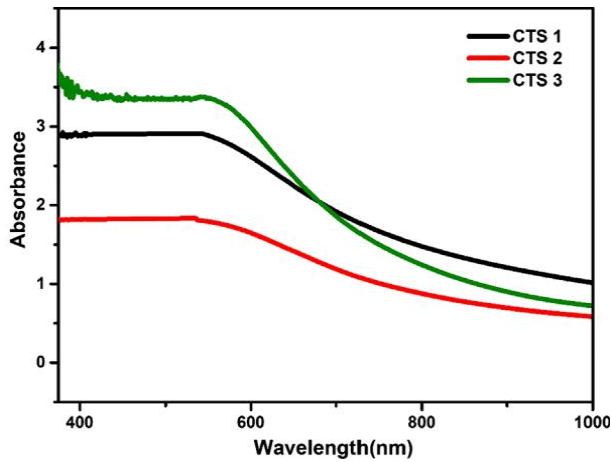

Fig. 4. Absorbance spectra of CTS thin films.

### 3.4. Electrical measurements

Electrical Resistivity and charge carrier density of CTS thin films were measured by the Vander Pau and Hall Effect measurement technique. Hot probe analysis was performed for majority carrier identification and all the films have confirmed a p-type semiconductor nature. Carrier concentration, mobility and resistance are influenced by the defects in the thin films. As carrier charge density is varied the resistivity also is changed. The estimated values of resistivity and charge carrier density are listed in Table 3. Varying the tin and sulphur content we can vary the carrier charge density and resistivity of the thin films [18]. The formation of CuS will result in an enhanced hole charge carrier concentration in the films favouring an increase of charge carrier density. When Sn is added it leads formation SnS compound [19] which reduces the concentration of holes in the CTS lattice further results in the decrement of charge carrier density and to enhancement in the resistivity of the films.

### 3.5. Harmonic generations

The experimental set up used to evaluate SHG and THG efficiency of CTS films is presented in Fig. 6. The films were preliminary photo treated using a 532 nm continuous wave (100 mJ) laser modulated beam with frequency 50 HZ to avoid overheating which favour an enhancement of the NLO responses due to the additional formation of the photo polarized states which effectively interacts with the phonon subsystem which leads to enhanced nonlinear optical effects [20,21].

As a fundamental laser source, we have used Nd: YAG Quanta-Ray Indi nanosecond laser. The laser generates pulses with a wavelength of 1064 nm, duration 2.5–10 ns and energy up to 500 mJ. The Thorlabs polarizers with polarization degree about 99.997% and the Thorlabs interference filters at 355 nm and 532 nm with bandwidth 5 nm have been applied for varying the input energy and spectral range of the light. Two additional interference filters with 340 nm, 360 nm, 540 nm and 560 nm have been used to separate the fluorescent and background scattering parasitic background. The registration has been performed by HAMAMATSU photomultipliers with 1 ns relaxation interference filters. Generally, the method is similar to the methods described in our previous works [22].

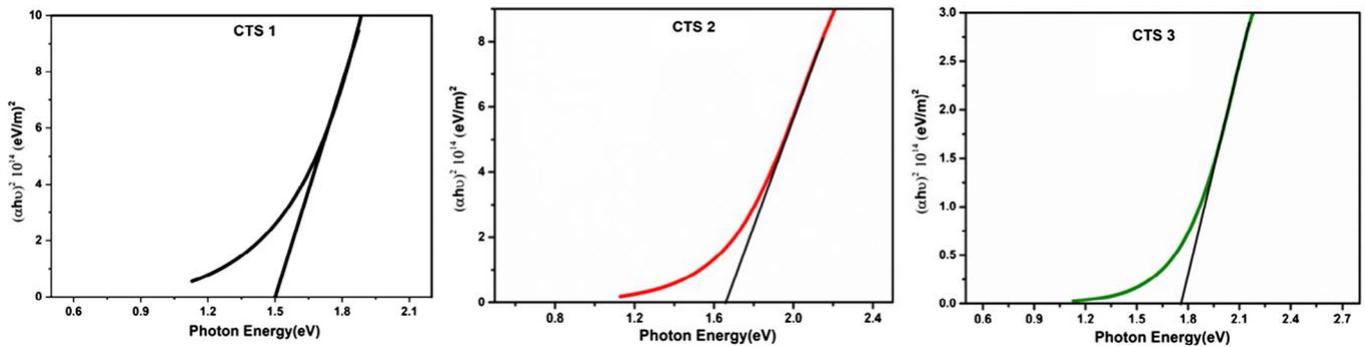

Fig. 5. Tauc absorption plot of CTS thin film.

Table 3
Electrical properties of CTS thin films.

| Sample | Resistivity (Ω-cm) | Carrier charge density × $10^{17}$ ($Cm^{-3}$) |
|---|---|---|
| CTS 1 | 11.1 | 14.9 |
| CTS 2 | 12.9 | 7.29 |
| CTS 3 | 25.3 | 3.71 |

The rotating table which allows achieving the angle's variation with respect to the sample surfaces giving the maximal signals (SHG or THG) was used for monitoring of the SHG/THG maxima. For the THG/SHG the interference filters with 355 nm and 532 nm nd spectral band widths 5 nm have been used. During the photo treatment, the photo-multiplier connected with the Tektronix oscilloscope has been detected the saturation and stabilization of the THG/SHG signal. The photo-inducing signal was stabilized by the intensity and the measurements of the SHG/THG have been performed with respect to the reference $BiB_3O_6$:Nd crystal crystals and lead-based optically active glasses [23,24]. The maximal of THG/SHG signals have been detected by manual rotation of the samples in the 3 axes by observation of the maximal nonlinear optical signals at some angle. Generally, the de-termination of the corresponding NLO susceptibilities was similar to the described in the ref. [25].

Figs. 7 and 8 represents the dependence of the SHG and THG versus the photoinduced beams. It is clear that all the nonlinear optical effi-ciencies (THG and SHG) signal increase with the enhancement of the fundamental energy density beams. For the SHG the maximal second-order susceptibility values have been obtained for the CTS 2. For the case of the THG maximal nonlinear optical susceptibilities was achieved for the CTS3. Such difference may be explained by different mechan-isms of the laser-stimulated NLO response of SHG and THG.

The achieved photo induced second order optical susceptibility at 1064 nm for the CTS 2 film was equal to about 0.89 pm/V and the minimal second order NLO susceptibility was achieved for the CTS 3 film (about 0.22 pm/V) while for the THG, the maximum efficiency is for the CTS 3 film. Such differences may be caused by a fact that the SHG is determined prevailingly by local charge density acentricity and for the THG it is defined by the differences of the excited and ground state dipole moments. Without preliminary 532 nm CW laser treatment photo induction, the signals are observed to be at least 20% less. The photo induced increase of temperature did not exceed 7 K which do not principally change the observed NLO effects in the films. Generally the laser stimulated SHG and THG has a power dependence slightly higher than for the traditional nonlinear optical effects which may be explained by preliminary phototreatment. Such treatment cause an occurrence of additional internal dc-electric field and the doping by the Sn and S causes an additional enhancement of the hyperpolarizabilites both due to electron as well as ionic photo polarization. As a consequence we have an increase of initial NLO susceptibilities. It is also noted that the present materials have a laser energy density threshold equal to about 210 J/m$^2$.

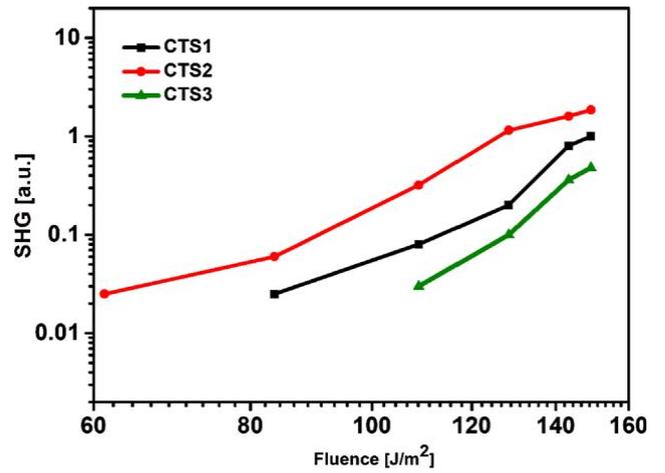

Fig. 7. Dependence of the SHG versus the fundamental energy density in the regime of the photo induced laser treatment.

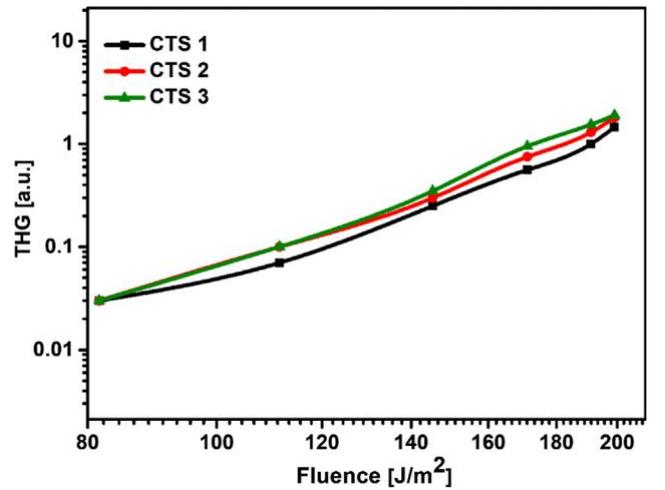

Fig. 8. Dependence of the THG versus the fundamental energy density in the regime of the laser stimulated treatment.

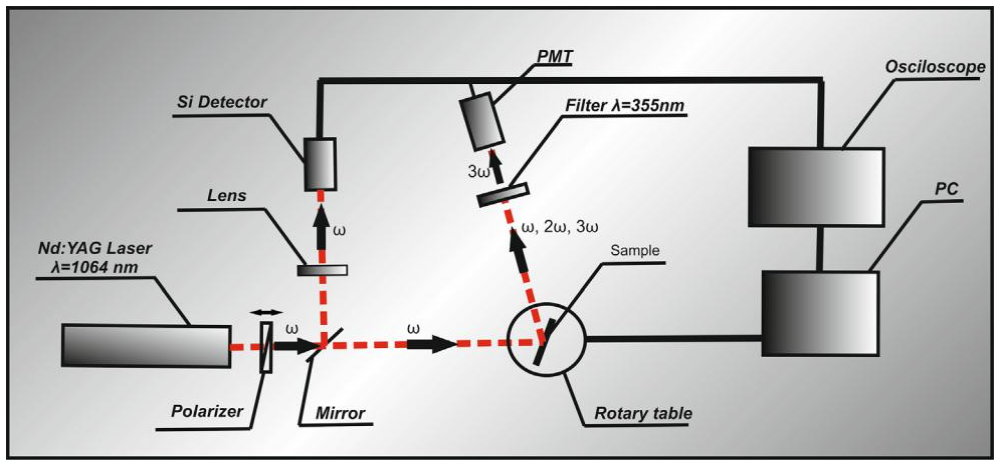

Fig. 6. Experimental set up of harmonic generation measurement.

4. Conclusions

Solution-based synthesis of CTS thin films for the nonlinear optical applications was reported. The grown films show a tetragonal crystal structure as evident from XRD studies and confirmed using Raman spectroscopy. The obtained Raman spectra were assigned to phonon mode as per zone centre phonon representation of optical and acoustic modes. The CTS film shows two Raman peaks which are identified to dominant "A" symmetry modes. The morphological studies indicate that formed grains are densely packed with well-defined shape and surfaces are free of pinholes. The maximally achieved photo induced second order optical susceptibility for the CTS 2 was equal to about 0.89 pm/V and the minimal second order NLO susceptibility was achieved for the CTS 3 film (about 0.22 pm/V). While for the THG the maximum efficiency is for the CTS 3 film. The electrical analysis shows that when Sn is added leading to the formation SnS compound which reduces the concentration of holes in the CTS lattice.

Acknowledgement

The presented results are part of a project that has received funding from the European Union's Horizon 2020 research and innovation program under the Marie Skłodowska-Curie grant agreement No 778156. I.V.K.,K.O.,J.J., acknowledge support from resources for sci-ence in the years 2018-2022 granted for the realization of international co-financed project Nr W13/H2020/2018 (Dec. MNiSW 3871/H2020/ 2018/2).